\documentclass[]{aastex631}

\usepackage{comment}
\usepackage{color}

\newcommand{\Hii}{H{$\,${\sc ii}}} 
\shorttitle{GGD12-15}
\shortauthors{Shimoikura et al.}
\graphicspath{{./}{figures/}}

\begin{document}

\title{Cluster Formation in GGD12-15: Infall Motion with Rotation of the Natal Clump}

\correspondingauthor{Tomomi Shimoikura}
\email{ikura@otsuma.ac.jp}

\author[0000-0002-1054-3004]{Tomomi Shimoikura}
\affiliation{Otsuma Women's University \\
 Chiyoda-ku, Tokyo, 102-8357, Japan}

\author[0000-0001-8058-8577]{Kazuhito Dobashi}
\affiliation{Tokyo Gakugei University \\
Koganei, 184-8501, Tokyo, Japan}

\author[0000-0001-9304-7884]{Naomi Hirano}
\affiliation{Academia Sinica Institute of Astronomy and Astrophysics,11F of Astronomy-Mathematics Building, AS/NTU, No.1, Sec. 4, Roosevelt Rd, Taipei, 10617, Taiwan}

\author[0000-0001-5431-2294]{Fumitaka Nakamura}
\affiliation{National Astronomical Observatory of Japan, Mitaka, Tokyo, 181-8588, Japan}
\affiliation{Department of Astronomical Science, School of Physical Science, SOKENDAI (The Graduate University for Advanced Studies), Osawa, Mitaka, Tokyo, 181-8588, Japan}

\author[0000-0003-1659-095X]{Tomoya Hirota}
\affiliation{National Astronomical Observatory of Japan, Mitaka, Tokyo, 181-8588, Japan}
\affiliation{Department of Astronomical Science, School of Physical Science, SOKENDAI (The Graduate University for Advanced Studies), Osawa, Mitaka, Tokyo, 181-8588, Japan}

\author[0000-0002-8125-4509]{Tomoaki Matsumoto}
\affiliation{Faculty of Sustainability Studies, Hosei University, Chiyoda-ku, Tokyo, 102-8160, Japan}

\author[0000-0003-4402-6475]{Kotomi Taniguchi}
\affiliation{National Astronomical Observatory of Japan, Mitaka, Tokyo, 181-8588, Japan}

\author[0000-0001-9368-3143]{Yoshito Shimajiri}
\affiliation{National Astronomical Observatory of Japan, Mitaka, Tokyo, 181-8588, Japan}




\begin{abstract}
We report results of observations of the GGD12-15 region, where cluster formation is actively taking place, using various molecular emission lines.
The C$^{18}$O $(J=1-0)$ emission line reveals a massive clump in the region with a mass of $\sim2800 M_\sun$ distributed over  $\sim2$ pc. 
The distribution of the C$^{18}$O $(J=3-2)$ emission is similar to that of a star cluster forming therein, with an elliptical shape of $\sim1$ pc in size. 
A bipolar molecular outflow driven by IRS 9Mc, a constituent star of the cluster, is blowing in a direction perpendicular to the elongated C$^{18}$O $(J=3-2)$ distribution, covering the entire clump. 
There is a massive core with a radius of 0.3 pc and a mass of $530 M_\sun$ in the center of the clump.
There are two velocity components around the core, which are prominent in a position-velocity (PV) diagram along the major axis of the clump.
In addition, a PV diagram along the minor axis of the clump, which is parallel to the outflow, shows a velocity gradient opposite to that of the outflow. The velocity structure strongly indicates the infalling motion of the clump.
Comparison of the observational data with a  simple model of infalling oblate clumps indicates that the clump is undergoing gravitational contraction with rotation.

\end{abstract}




\section{Introduction} \label{sec:intro}
The majority of stars in our Galaxy are formed in clusters \cite[e.g.,][]{ladalada,Ridge2003,Shimoikura2018}, but it is still unclear how cluster formation is initiated. 
To study the evolutionary process of cluster formation, it is necessary to observe molecular clumps and investigate the dynamics of gas in an early stages of evolution.

Revealing how mass accumulates from a parent molecular clump to form a massive core is important for our understanding the formation process of clusters.
In regions such as NGC 2264 and Serpens South which are thought to be in an early stage of cluster formation, mass accretion onto massive cores from the surroundings has been reported \citep[e.g.,][]{ Peretto2006, Kirk2013}.
\cite{Peretto2013} suggested that in a massive infrared dark cloud SDC335, the global collapse of the surrounding cloud causes a large amount of mass to accumulate in its center.
By analyzing gas kinematics of massive clumps with a mass of $\sim10^{3}M_\sun$, we found in our previous statistical studies that some molecular clumps are characterized by a velocity structure  representing gravitational contraction with rotation \citep{Shimoikura2016, Shimoikura2018}. 
We suggested that such infall motion with rotation of the clumps is observed only in an early stage of cluster formation.

Some other reports have identified candidates of infalling massive clumps by examining asymmetric profiles of optically thick molecular emission lines \citep[e.g.,][]{Reiter2011, He2016}.
Cluster-formation can also be triggered by cloud-cloud or filament-filament collisions \citep[e.g.,][]{Nakamura_Ser, Dobashi2014, Dobashi_DR21}, and can also be influenced by supernova remnants \citep[e.g.,][]{Tatematsu1990, Dobashi_NCS}.  Other than these, there can also be various gas motions in cluster forming regions such as powerful outflows as well as inflows caused by multiple protostars. Therefore, in order to understand the gas kinetics of cluster-forming clumps through molecular spectroscopy, it is necessary to investigate the entire velocity field of cluster-forming clumps.

GGD12–15 is an active star-forming site embedded in the Monoceros R2 (Mon R2) molecular cloud \citep[e.g.,][]{Gyulbudaghian,Carpenter2008}.
It is listed as No.4978 in the near-infrared dense core catalog compiled by \cite{Dobashi2005} and \cite{Dobashi2011}.
The distance to the region is estimated to be 893 pc based on the Very Long Baseline Array observations \citep{Dzib2016}.
\cite{Heaton} showed a molecular clump with a mass of $\sim10^{3}M_\sun$ in this region based on their $^{13}$CO($J=1-0$) observations. We will hereafter refer to the clump in this region as the GGD$12–15$ clump. 

A compact {\Hii} region VLA1 excited by a B0.5 star is formed near the center of the clump \citep{Gomez}.
There is also a bipolar molecular outflow in the clump extending about $6\arcmin$ in the southeast-northwest direction \citep{Rodriguez1982,Little,Qin}, 
which seems to be driven by a mid-infrared source IRS 9Mc located about $0.5\arcmin$ northeast of VLA1 \citep{Sato}.
\cite{Sato} suggested that IRS 9Mc is a Class I/0 object.
Moreover, a massive protostar consistent with an early-B spectral type has been identified near VLA1 \citep{Maaskant}.

The GGD12-15 clump is accompanied by a young cluster with $98\pm10$ members \citep{Gutermuth2005,Gutermuth2011}. 
Many young stellar objects (YSOs) have been identified, 
and their distribution shows that Class I objects are concentrated around the center of the clump, while Class II objects are widely spread over the region \citep{Gutermuth2011}. 
\cite{Maaskant} estimated the median age of the cluster to be $\sim4$ Myr, and that of  VLA1 to be $<1$ Myr, 
for which they suggested sequential star formation from intermediate-mass to massive stars along the line of sight in this region.

These results indicate that the GGD12-15 clump is forming the cluster very actively.
However, the region observed in previous studies is limited, and the overall spatial and velocity structures of the clump are not clear yet.  Therefore, the relationship between the cluster and the clump has not been understood well. 
In addition, the region has been observed using only limited molecular emission lines such as CO.
It is not evident if there are dense cores embedded in the parent molecular clump.
To investigate the molecular gas kinematics in the clump, we observed the region with various molecular emission lines using the Nobeyama 45-m telescope and the 15-m James Clerk Maxwell Telescope (JCMT).

This paper is organized as follows.
In Section \ref{sec:obs}, we describe the observations. 
We identified the GGD12-15 clump extending over $\sim2$ pc, 
and its physical parameters are derived in Section \ref{sec:results}.
The detailed velocity structure of the outflow associated with the clump is also investigated.
In Section \ref{sec:discussion}, we analyze the velocity structures of the clump,
and discuss the collapsing motion with rotation of the clump.
We present the summary of this paper in Section \ref{sec:conclusions}.


\section{Observations} \label{sec:obs}

\subsection{Observations with the NRO 45-m Telescope} \label{sec:obs1}
Observations toward the GGD12-15 region in the 100 GHz band were carried out in March 2017 and March 2019 using the 45 m telescope at the Nobeyama Radio Observatory (NRO). 
The half-power beam width (HPBW) of the telescope is $\sim15\arcsec$ at 110 GHz.
We performed mapping observations of an area of $10\arcmin$ around GGD12-15 in the On-The-Fly mode \citep{Sawada}. 
We used the four-beam receiver \citep[FOREST;][]{Minamidani} as the frontend and the digital spectrometer `SAM45' as the backend.
The spectrometer was set to a bandwidth of 31.25 MHz and a channel width of 15.26 kHz ($\sim0.04$ km s$^{-1}$), and the Spectral Window mode was used.
Five molecular emission lines, $^{12}$CO($J=1-0$), $^{13}$CO($J=1-0$), C$^{18}$O($J=1-0$), CCS ($J_N=8_7-7_6$), and N$_2$H$^{+}$($J=1-0$), were observed in 2017. 
Additional three molecular lines, SO($J_N=3_2-2_1$), CS ($J=2-1$), and HC$_3$N ($J=10-9$), were taken for the same area in 2019. 

The SiO maser source Gx Mon was used as a pointing source and observed every 1.5 hours, 
confirming that the pointing accuracy was better than $\sim5\arcsec$ each time.
The intensity was calibrated using the chopper wheel method \citep{Kutner}. We observed another star-forming region S235AB \citep{Shimoikura2016} every observing day to correct the intensity variation.

The data were calibrated using `NOSTAR', a reduction software package of NRO \citep{Sawada}.
The obtained antenna temperature $T_{\rm a}^{*}$ data were converted to the brightness temperature $T_{\rm mb}$ using a scale factor based on beam efficiencies ($\eta_{\rm mb}$) measured annually by NRO.

\subsection{Observations with JCMT} \label{sec:obs2}

Observations toward the GGD12-15 region in the 300 GHz band were carried out in December 2018 
using JCMT at Mauna Kea in Hawaii. 
HPBW of the telescope is $\sim14\arcsec$ at 350 GHz.
The observations were allocated as the program No.M18BP015.
We performed mapping observations of the same area observed with the 45 m telescope in Raster mode. 
We used the Heterodyne Array Receiver Program \citep[`HARP' ;][]{Buckle} as the frontend and the digital spectrometer `ACSIS' as the backend.
The spectrometer was set to a bandwidth of 250 MHz and a channel width of 30.5 kHz ($\sim0.03$ km s$^{-1}$).
Two molecular emission lines of $^{13}$CO($J=3-2$) and C$^{18}$O($J=3-2$) were observed.

Using a standard JCMT process of calibration observations including pointing, focus, and standard source observations were performed during the observation period.
The data were reduced using `Starlink', a reduction software package of JCMT observations.
$\eta_{\rm mb}$ of 0.64 was used following the JCMT website to convert the obtained data to $T_{\rm mb}$.\\

All of data from both observations were re-gridded onto common cubes with $7\farcs5$ pixels and smoothed with a velocity resolution of 0.1 km s$^{-1}$. 
The resulting 1 $\sigma$ noise is in the range $\Delta T_{\rm mb}=0.2 - 0.7$ K.
The spatial resolution of the final maps is $22\arcsec-24\arcsec$.
The observed molecular lines and the noise levels for each line are summarized in Table \ref{tab:line}.

\section{Results} \label{sec:results}

\subsection{Spatial distribution of molecular emission lines} \label{sec:ii}
\begin{figure}
\begin{center}
\includegraphics[scale=0.4]{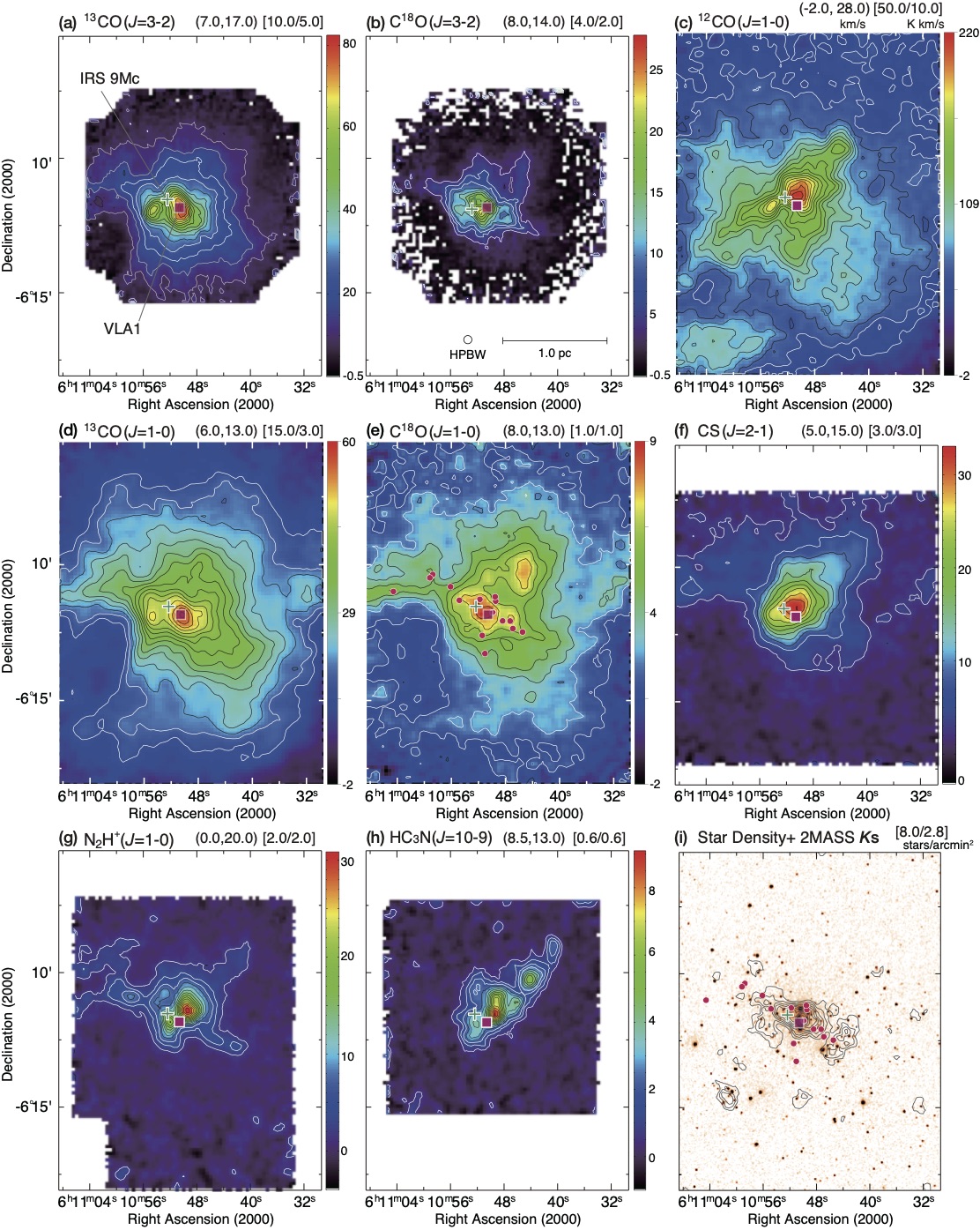}
\caption{
(a)--(h) Maps of the integrated intensity of the observed emission lines around GGD12-15. 
Molecules, transitions, and the velocity ranges in units of km s$^{-1}$ used for the integration are indicated in the parentheses in each panel. The lowest contours and the contour intervals in units of K km s$^{-1}$ are indicated in the square brackets above each panel. 
(i) Map of the star density overlaid with the 2MASS $K_{\rm s}$ image (red scale). 
The lowest contour and the contour interval in units of arcmin$^{-2}$ are indicated in the panel. 
The positions of the compact H{$\,${\sc ii}} region VLA1 \citep{Gomez} and the powering source of molecular outflow IRS 9Mc \citep{Sato} are marked in all of the panels.
The Class 0/I objects classified by \cite{Gutermuth2011} are shown as red circles in panels (e) and (i).
\label{fig:ii}}
\end{center}
\end{figure}

We show in Figure \ref{fig:ii} a series of the integrated intensity maps of the observed molecular lines
\footnote{
The integral intensity maps of SO and CCS that are not directly used for the analysis in this study are given in Appendix \ref{sec:appendix}.}.
The positions of the compact H{$\,${\sc ii}} region VLA1 \citep{Gomez} and the powering source of molecular outflow IRS 9Mc \citep{Sato} are marked in all of the panels.
The GGD12-15 clump is distributed over $\sim2$ pc, and the spatial structures appear different depending on the observed molecular lines.
The clump has a thin arm-like structure on its western and eastern sides, which is well seen in the C$^{18}$O $(J=1-0)$ map. 
According to the large map of the region \citep{Gyulbudaghian, Dobashi2011,Pokhrel2016}, this extends from the star-forming region Mon R2, indicating that the clump is a part of the region.

In the last panel of Figure \ref{fig:ii}, we also show the distribution of a star density around the observed area created using the 2MASS Point Source Catalog \citep{Skrutskie}. 
The distribution of the star density reveals the spatial structure of the cluster in this region.
The distribution of the Class 0/I objects \citep{Gutermuth2011} is superimposed on the C$^{18}$O $(J=1-0)$ map in Figure \ref{fig:ii}(e).
The cluster extends from northeast to southwest, and the Class 0/I objects are located along with the structure.
The size of the cluster is found to be $\sim1$ pc and the number of stars is $\sim90$, as determined by the same method as \cite{Shimoikura2018}.
The properties are consistent with that reported by \cite{Gutermuth2005}.

The spatial distribution of the $^{13}$CO, C$^{18}$O, and CS emission lines have an elongated structure from northeast to southwest, similar to the cluster's spatial structure. 
In particular, the distribution of C$^{18}$O ($J=3-2$) shows an elliptical structure, which correlates well with the extent of the cluster.
We also found that the strongest peak position in the observed intensity maps is different for each molecular emission line, and that of C$^{18}$O $(J=3-2)$ agrees well with the position of  VLA1.
This implies that the C$^{18}$O $(J=3-2)$ line traces a relatively high temperature region of the clump.
On the other hand, the position of the intensity peak of CS, N$_2$H$^{+}$, and HC$_3$N do not coincide with VLA1, and each emission appears to extend along the southeast-northwest direction, which is perpendicular to the cluster distribution.

\begin{figure}
\begin{center}
\includegraphics[scale=.4]{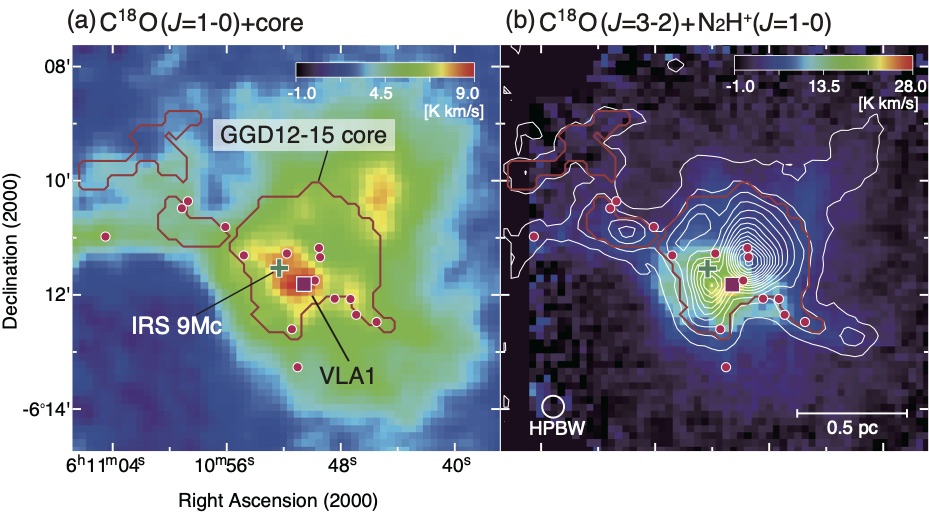}
\caption{
Distribution of the identified cores (the red contours) overlaid on (a) the C$^{18}$O$(J=1-0)$ map and (b) the maps of N$_2$H$^{+}$ (the white contours) and C$^{18}$O$(J=3-2)$ (color scale).
We refer to the largest core of the cores as the GGD12-15 core. 
The symbols are the same as in Figure \ref{fig:ii}, with the purple square for VLA1, the green plus sign for IRS 9Mc, and the red circle for the Class 0/I objects.
\label{fig:ii_large}}
\end{center}
\end{figure}

We compared the observed maps with the $850 \micron$ dust emission in the GGD12-15 region obtained with JCMT \citep{Gutermuth2005}.
The dust emission corresponds well to the structure of the N$_2$H$^{+}$ emission.
Based on the N$_2$H$^{+}$ map, we defined cores within the clump as the areas extending larger than the observed beam. 
We applied the dendrogram algorithm \citep{Rosolowsky} to the N$_2$H$^{+}$ map.
The minimum threshold intensity required to identify a parent tree structure and
the splitting threshold intensity required to identify structures were set to $5 \sigma$, respectively.
We used the $astrodendro$ Python package\footnote{http://www.dendrograms.org/} to compute dendrograms and set parameters to min$\_$value$=1.0$ K km s$^{-1}$, min$\_$delta$=1.0$ K km s$^{-1}$, and min$\_$npi$=25$ pixels which is equivalent to a synthesized beam of 0.39 arcmin$^2$).
We used the results of dendrogram leaves as `cores'.
As a result, three cores were identified, and the cluster formation was found to occur in the largest core located in the center of the clump. 
Hereafter, we will refer to the largest core as the GGD12-15 core.
The core is shown in Figure \ref{fig:ii_large}, together with 
the intensity maps of C$^{18}$O$(J=1-0)$ and C$^{18}$O$(J=3-2)$.

It should be noted that the C$^{18}$O($J=1-0$) map shows a hole-like structure on the northwest side of the core whose peaks are evident in N$_2$H$^{+}$.
The result implies that the C$^{18}$O molecule is adsorbed onto dust grains in the dense and cold part of the clump. The  N$_2$H$^{+}$ ion can survive in the part owing to the depletion of the CO molecule there \cite[e.g.,][]{Bergin2002, Taniguchi2019}.
The N$_2$H$^{+}$ map also shows a hole toward the peak of C$^{18}$O($J=3-2$). 
This can be interpreted that N$_2$H$^{+}$ is destructed due to the high temperature \cite[e.g.,][]{Lee2004}. 

\subsection{Physical properties of the clump and the core} \label{sec:mass}
 Under the assumption of the local thermodynamic equilibrium (LTE), we used the C$^{18}$O ($J=1-0$) data to derive the mass of the entire clump.
The excitation temperature of the molecular emission line, $T_{\rm ex}$, was determined by assuming that the $^{12}$CO molecule is optically thick and using its $T_{\rm mb}$, by the following equation,
\begin{equation}
\label{eq:tex}
T_{\rm ex}=5.53\left\{ \ln\left[1+\frac{5.53}{T_{\rm mb}+0.819}\right]\right\}^{-1}.
\end{equation}
The maximum value of $T_{\rm ex}$ is 40.5 K, which coincides with the position of VLA1.

The column density of each molecule $X$ (e.g., $X$=C$^{18}$O), $N(X)$, was calculated from the following equation \cite[e.g.,][]{Mangum2015},
\begin{equation}
\label{eq:column}
N(X)=\frac{3h}{8\pi^{3}}\frac{Q}{\mu^{2}S_{ij}}\frac{e^{Eu/k{T}_{\mathrm{ex}}}}{e^{h\nu/kT_{\mathrm{ex}}}-1}\int \tau dv,
\end{equation}
where 
$k$, $h$, and $\nu$ are the Boltzmann constant, the Planck constant, and the rest frequency of the observed emission line, respectively.
$Q$ is the partition function approximated as $Q=k\,T_{\rm ex}/h\,B_{0}+1/3$ \citep[e.g.,][]{Mangum2015} where $B_{0}$ is the rotational constant of the molecule. 
$\mu$ is the dipole moment, $E_{u}$ is the energy of the upper level, $S_{ij}$ is the intrinsic line strength of the transition for $i$ to $j$ state, and $\tau$ is the optical depth.
For the estimation, we used the spectral line parameters taken from Splatalogue\footnote{https://splatalogue.online/}.

The column density of C$^{18}$O derived from the $J=1-0$ transition, $N$(C$^{18}$O), was calculated using Equation (\ref{eq:column}).
To convert $N$(C$^{18}$O) to molecular hydrogen column density $N$(H$_2$) for each pixel, 
we assumed $N$(C$^{18}$O)\,$ =1.7\times10^{-7}N(\rm H_{2})$ \citep{Frerking}.
The mass, $M_{\rm LTE}$, of the clump was calculated using the following equation.
\begin{equation}
\label{eq:mass}
{M_{\rm LTE}} = \overline{\mu}{m_{\rm H}}\int_S {N({\rm{H_2}})dS},
\end{equation} 
where $m_{\rm H}$ is the hydrogen mass, and $\overline{\mu}$ is the mean molecular weight, 
which is assumed to be 2.8. 
$S$ is the surface area of the clump defined as the sum of the regions surrounded by the contour with the integrated intensity of C$^{18}$O greater than $5 \sigma$, and is $\sim80$ arcmin$^2$. 
We estimated $M_{\rm LTE}$ of the clump to be 2850 $M_\sun$.
We also estimated the mass integrated over the region enclosed by the contour at $50\%$ of the peak intensity of the clump
($\sim12$ arcmin$^2$) to be 1010 $M_\sun$, and the mass integrated over the entire region shown in Figure \ref{fig:ii} (f) ($\sim117$ arcmin$^2$) to be 3230 $M_\sun$.
\cite{Pokhrel2016} estimated the mass of the GGD12-15 region (their Region 4) to be 1773 $M_\sun$ based on the $Herschel$ data.
The difference between the results of \cite{Pokhrel2016} may be due to differences in the regions $S$.

We also derived the virial mass, $M_{\rm vir}$, as
\begin{equation}
\label{eq:M_Vir}
M_{\rm vir}=
\frac{5R}{G}
\frac{\overline{\Delta V}^{2}}{8\,{\rm ln}2}\ ,
\end{equation} 
where $G$ is the gravitational constant, $R$ is the radius of the clump derived as $\sqrt{S/\pi}$, and 
$\overline{\Delta V}$ is the mean line width (FWHM)
in the composite profile obtained by averaging all of the spectra within $S$. 
$R$ and $\overline{\Delta V}$ were estimated to be 1.3 pc and 2.0 km s$^{-1}$, respectively.
We also derived the LSR velocity $V_{\rm{LSR}}=11.4$ km s$^{-1}$ 
and the line width $\Delta V=2.5$ km s$^{-1}$ at the peak intensity position of the clump
by fitting the spectra with a single Gaussian function. 

To derive mass of the GGD12-15 core, we used the N$_2$H$^{+}$ data.
We applied the hyperfine spectra fitting to the data in the same way as \cite{Shimoikura2019}. 
We then derived the parameters such as $T_{\rm ex}$,
$V_{\rm{LSR}}$, ${\Delta V}$, $T_{\rm mb}$,
and the total optical depth $\tau_{\rm{tot}}$ for all of the hyperfine components of each pixel. 
The column density of N$_2$H$^{+}$, $N$(N$_2$H$^{+}$), was derived as the following formula \cite[e.g.,][]{Caselli2002,Mangum2015},
\begin{equation}
\label{eq:column2}
N({\rm N}_{2}{\rm H}^{+})=\frac{8\pi^{3/2}}{2\sqrt{\rm{ln}2}}\frac{Q {\nu^{3}}}{c^{3}A g_{u}}\frac{\tau_{\rm tot}\Delta V}{1-e^{-{h\nu/kT_{\rm ex}}}} \ , 
\end{equation}
where $\nu$ is the rest frequency of N$_2$H$^{+} (J=1-0)$. $g_{u}$ is the degeneracy of the upper level of a transition and  
$A$ is the Einstein coefficient for spontaneous emission, which we used 3 and $3.628 \times 10^{-5}$\citep{Pagani2009}, respectively.
$N(\rm H_{2})$ for the core was determined from $N$(N$_2$H$^{+}$)\,$ =3.0\times10^{-10}N(\rm H_{2})$\citep{Caselli2002}.
We finally estimated $M_{\rm LTE}$ of the core using Equation (\ref{eq:mass}).
$S$ of the core was defined by the area determined by the core identification (Figure \ref{fig:ii_large}).
The $M_{\rm LTE}$ of the core was estimated to be $530$ $M_\sun$.
$\overline{\Delta V}$, $R$, $M_{\rm vir}$ of the core were obtained in the same way as for the clump.
We summarize the physical properties obtained in this subsection in Tables \ref{tab:param1}$-$\ref{tab:mass}.

Here, as we pointed out in Section \ref{sec:ii}, there is an area where C$^{18}$O may be less abundant due to adsorption onto dust grains, and this may lead to an underestimation of $N$(H$_2$) for the clump mass.
For example, the mass of the same region as the GGD12-15 core is $\sim10\%$ smaller if we used $N$($\rm H_{2})$ estimated from the C$^{18}$O ($J=1-0$) line instead of N$_2$H$^{+}$.
We also note that if the amount of N$_2$H$^{+}$ decreases toward the center of the GGD12-15 core, the mass obtained from N$_2$H$^{+}$ may be a lower limit because of the destruction of N$_2$H$^{+}$ \citep[e.g.,][]{Lee2004}.
The mass estimate also depends on other factors such as $T_{\rm ex}$.
The clump mass was found to fluctuate by $\sim40\%$ if we assume a uniform $T_{\rm ex}$ using the minimum and maximum values of $T_{\rm ex}$ ($10-40$ K) throughout the clump. 
Taking into account these ambiguities, the LTE mass estimates are probably uncertain by $40\%$ at most.

We compare the derived values of $M_{\rm LTE}$ and $M_{\rm vir}$ of the GGD12-15 core and the clump.
The results for the core infer that $M_{\rm LTE}$ is greater than $M_{\rm vir}$ by a factor of 3, indicating that the core is gravitationally bound.
In the case of the entire clump, $M_{\rm LTE}$ is more than twice larger than $M_{\rm vir}$, indicating that the clump is gravitationally unstable as a whole.

\subsection{Search for Outflows} \label{sec:outflow}
\begin{figure}
\begin{center}
\includegraphics[scale=.4]{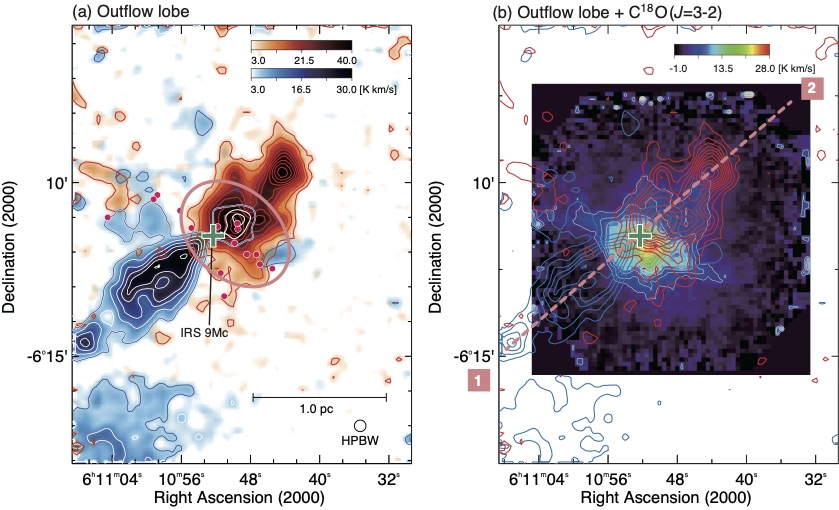}
\caption{
(a) The $^{12}$CO intensity maps for the blue and red lobes of the outflow found around the clump. 
The blue contours denote the blue lobe, and the red contours denote the red lobe. 
The lowest contours and the contour intervals for both lobes are 5 K km s$^{-1}$, respectively.
The pink ellipse indicates the overlap between the two lobes (see text).
(b) The outflow lobes overlaid on the C$^{18}$O ($J=3-2$) map (color scale).
The pink dashed line  (labeled $1-2$) indicates the positions used to create PV diagrams in Figure \ref{fig:outflow2}.
\label{fig:outflow1}}
\end{center}
\end{figure}

We searched for outflows in the observed region by examining the $^{12}$CO spectra in all pixels.
As a result, a blue-shifted molecular gas component was detected in the velocity range $-2\le V_{\rm LSR}\le6$ km s$^{-1}$, 
and a red-shifted molecular gas component was detected in the velocity range $15\le V_{\rm LSR}\le30$ km s$^{-1}$.
Figure \ref{fig:outflow1} (a) shows the integrated intensity maps for these velocity ranges.
The lobes are large, covering the entire clump.
The blue lobe extending to the southwest and the red lobe extending to the northeast centered on IRS 9Mc were identified as in \cite{Little}, and the blue lobe is found to be more extended than the map from previous studies.
We also found that the lobes have a structure in which the blue and red lobes partially overlap in the vicinity of the driving source, which is indicated by the pink ellipse in the figure.
This structure can be reproduced if the molecular flow axis is inclined about $30^\circ$ to the line of sight and the opening angle of the flow is about $60^\circ$ as shown by \cite{Cabrit} in their model.

In Figure \ref{fig:outflow1} (b), the C$^{18}$O $(J=3-2)$ map is overlaid on the distribution of the lobes. 
Each lobe extends perpendicular to the elongated distribution of the C$^{18}$O emission.
In particular, the red lobe appears to overlap with the southeast-northwest extending structure of the HC$_3$N emission line, 
suggesting that the molecule can be enhanced in the region shocked by the outflow \cite[e.g.,][]{Shimajiri2015,Taniguchi2018}.
To investigate whether the HC$_3$N emission is caused by the outflow, 
we searched for a high velocity component showing indications of the outflow in the spectra of the emission,
but could not detect any obvious components.

\begin{figure}
\begin{center}
\includegraphics[scale=.4]{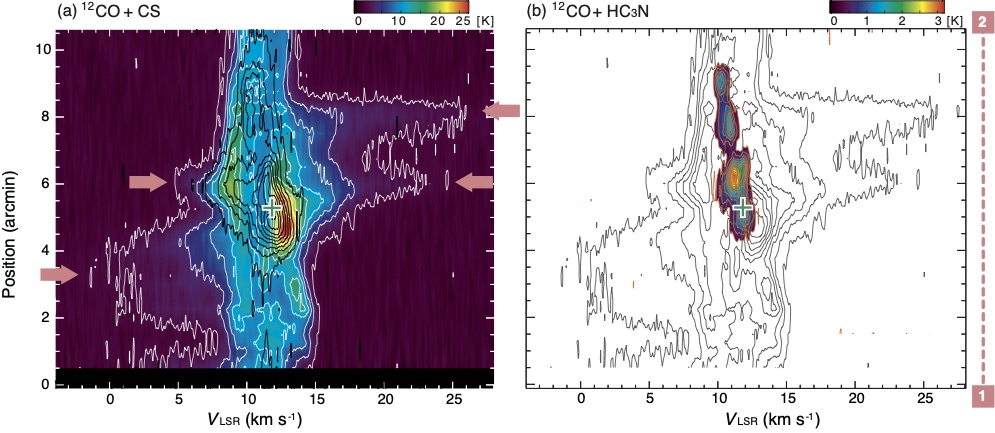}
\caption{
(a) Position–velocity diagrams of the $^{12}$CO (color scale and white contours) and CS (black contours) emission lines taken along the cut labeled 1-2 in Figure \ref{fig:outflow1}(b). 
The lowest contours and the contour intervals are 1.3 K and 3.0 K for the $^{12}$CO emission line.
The lowest contours and the contour intervals are 0.5 K and 1.0 K for the CS emission line.
The pink arrows indicate the molecular flow structure that appears to erupt in two stages on either side of the red and blue lobes (see text).
(b) Same as panel (a), but with the HC$_3$N emission line superimposed in color scale and orange contour.
The lowest contours and the contour intervals are 0.3 K for the HC$_3$N emission line.
The green plus denotes the position of IRS 9Mc.
\label{fig:outflow2}}
\end{center}
\end{figure}

Figure \ref{fig:outflow2} (a) displays a position-velocity (PV) diagram of the $^{12}$CO emission measured along cuts centered at the position of IRS 9Mc.
The position used for the measurements is shown in Figure \ref{fig:outflow1} (b).
We also show a PV diagram of the CS emission, which was measured at the same position as $^{12}$CO. 
The CS molecule is known to be often strongly detected in regions affected by shocks such as bipolar molecular outflows.
A part of the outflow can be seen from the PV diagram of CS.
We found a structure in Figure \ref{fig:outflow2} (a) that looks like a molecular flow erupting in two stages on both sides of the red and blue lobes, as indicated by the pink arrows.
This suggests that the molecular outflows exhibit intermittency.

Figure \ref{fig:outflow2} (b) shows the PV diagram measured with the HC$_3$N emission overlaid on that of $^{12}$CO. 
We found that there is a velocity gradient between local peaks seen in the emission. 
It is noteworthy that the direction of the velocity gradient in these peaks is in the opposite sense to that of the outflow, suggesting that the clump is undergoing gravitational contraction.


\subsection{Contribution of outflow to the clump turbulence} \label{sec:dis1}

The physical properties of the outflow were calculated in the same way as in \cite{Shimoikura2018}.
We assumed that the $^{13}$CO emission is optically thin and that the [$^{12}$CO]/[$^{13}$CO] abundance ratio is equal to the terrestrial value of 89.
Using the ratio of the integrated intensities of $^{12}$CO and $^{13}$CO, $I(^{12}$CO) and $ I(^{13}$CO), 
the optical depth of $^{12}$CO for the high-velocity component of the outflow, $\tau_{\mathrm{CO}}$, was obtained for the intensity peak position of the lobes from the following equation,
\begin{equation}
\label{eq:tau}
\frac{I(^{12}\mathrm{CO})}{I(^{13}\mathrm{CO})}=89 \times \frac{1-e^{-\tau_{\mathrm{CO}}}}{\tau_{\mathrm{CO}}}.
\end{equation} 

The column density of $^{12}$CO, $N(^{12}$CO), of the lobes was calculated from Equation (\ref{eq:column}) using the derived $\tau_{\mathrm{CO}}$ and $T_{\rm ex}$.
For the conversion from $N(^{12}$CO) to $N$(H$_2)$, we adopted the relation of $N$(H$_2)=1.4\times10^{4} N(^{12}$CO) \citep{Frerking}.
Using the obtained parameters, the total mass of the lobes, $M_{\rm lobe}$, was estimated with Equation (\ref{eq:mass}).
We defined the systemic velocity $V_{\rm sys}$ of the clump as 11.4 km s$^{-1}$ based on the composite C$^{18}$O ($J=1-0$) spectrum within the surface area of the lobes $S_{\rm lobe}$
which was defined as the area above the 5 $\sigma$ noise level.
The maximum velocity $V_{\rm max}$ of each lobe was determined at the velocity where the high velocity component was detected at the 3 $\sigma$ noise level.
The dynamical time scale $t_{\rm d}$ of each lobe was calculated by finding the maximum distance $R_{\rm lobe}$ from the driving source to the tip of each lobe, and dividing it by $\vert V_{\rm max}-V_{\rm sys}\vert$.
We also calculated the mechanical momentum $P_{\rm lobe}$ as $M_{\rm lobe}\vert V_{\rm max}-V_{\rm sys}\vert$.
Here, $\tau_{\mathrm{CO}}$ is the upper limit of the optical depth, and we derived the properties of the outflowing gas as the upper limits. 
Furthermore, we estimated the lower limits for $M_{\rm lobe}$, $P_{\rm lobe}$, and $E_{\rm lobe}$  by assuming that the outflowing components in the $^{12}$CO emission line are optically thin ($\tau\ll0$).
The obtained physical properties for the two lobes are summarized in Table \ref{tab:outflow}.

As a result of the calculations, the masses of the blue lobe and red lobe are estimated to be $11.7 M_\sun$ and $10.6 M_\sun$, respectively. 
$t_{\rm d}$ is estimated to be $8.9\times 10^{4}$ yr for the blue lobe and $4.5\times 10^{4}$ yr for the red lobe, which is consistent with the values estimated by \cite{Rodriguez1982} ($3\times 10^{4}$ yr) and \cite{Little}  ($9\times 10^{4}$ yr).
The obtained $t_{\rm d}$ is also consistent with the age of the driving source IRS 9Mc (= the Class 0/I object), as already pointed out by \cite{Sato}.

We further estimated how much the detected outflow contributes to the turbulence of the entire clump.
Turbulence plays a role to support molecular clumps against self-gravity, and molecular outflows are considered to be the source of turbulence in molecular clumps \cite[e.g.,][]{Snell}.
The momentum injection rate by outflows into the clump, $\dot{P}_{\rm lobe}$, was obtained by summing $P_{\rm lobe}$ of the blue and red lobes and dividing it by the average value of $t_{\rm d}$ of the two lobes $\overline{t_{\rm d}}$.
We derived $\dot{P}_{\rm lobe}=6.1\times10^{-3} M_\sun$ km s$^{-1}$ yr$^{-1}$.
On the other hand, 
turbulence dissipation rate of the clump, $\dot{P}_{\rm turb}$, can be obtained by the following equation \citep{Nakamura2014}, 
\begin{equation}
\label{eq:Pturb}
\dot{P}_{\rm turb}=-0.21M_{\rm LTE}\,{\sigma_{\rm3D}^2}/R,
\end{equation} 
where $\sigma_{\rm3D}$ is the velocity dispersion, obtained as $\sqrt{3}(\overline\Delta V/\sqrt{8{\rm ln}2})$ using the clump parameters (Table \ref{tab:param2}).
We derived $\dot{P}_{\rm turb}=1.2\times10^{-3}M_\sun$ km s$^{-1}$ yr$^{-1}$.

Comparing  $\dot{P}_{\rm lobe}$ and $\dot{P}_{\rm turb}$, we found $\dot{P}_{\rm lobe}>\dot{P}_{\rm turb}$.
This value of $\dot{P}_{\rm lobe}$ is enough to maintain the turbulence of the parent molecular clump, 
but the clump would disperse if all of the $\dot{P}_{\rm lobe}$ is absorbed by the clump.
However, as we will discuss in Section 4, we found that the clump is gravitationally bound and collapsing. Therefore, we suggest that the outflow is injecting a marginal amount of turbulence into the natal clump and has little effect on the clump dynamics.


\subsection{Velocity Structure of the Clump} \label{sec:discussion}

\begin{figure}
\begin{center}
\includegraphics[scale=.4]{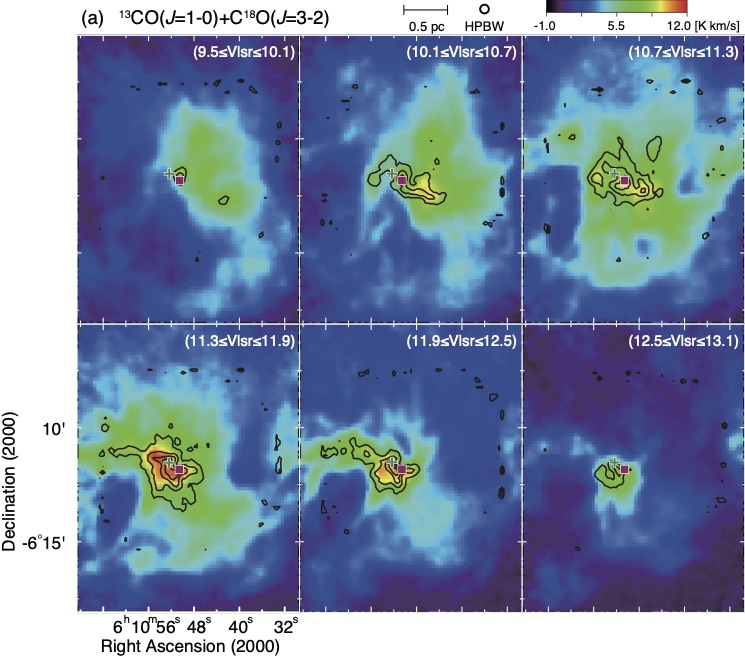}
\caption{
Velocity channel maps of the $^{13}$CO$(J=1-0)$ and C$^{18}$O$(J=3-2)$ emission lines 
made at every 0.6 km s$^{-1}$ in the velocity range $9.5\leqq V_{\rm LSR}\leqq 13.1$ km s$^{-1}$.
The channel maps for the $^{13}$CO emission line are shown in color scale, and that of the C$^{18}$O emission line are shown by the black contours.
The lowest contour and contour interval are 1.4 K km s$^{-1}$, respectively.
\label{fig:channel}}
\end{center}
\end{figure}
\begin{figure}
\begin{center}
\includegraphics[scale=.4]{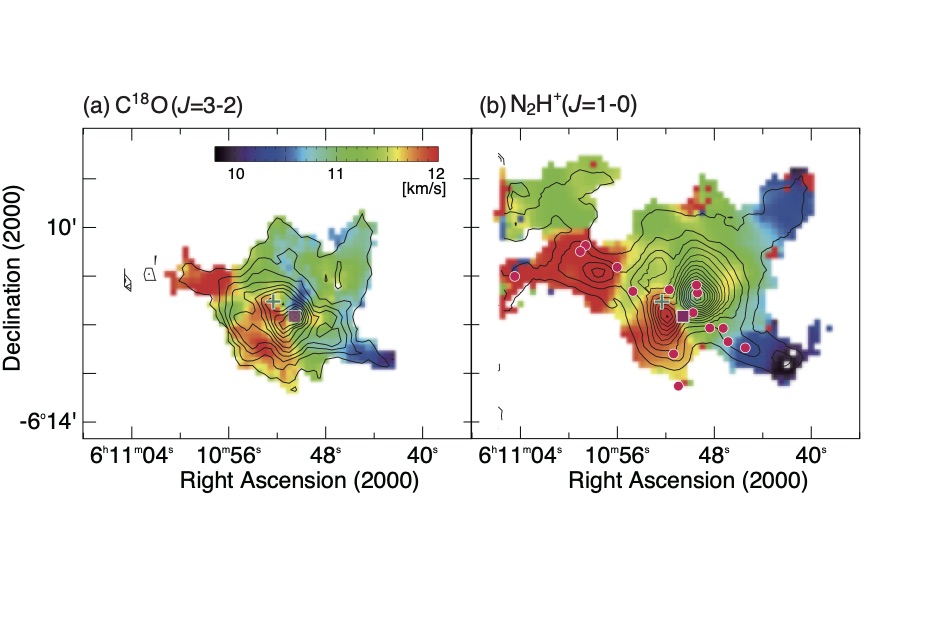}
\caption{
Intensity-weighted mean velocity map over the velocity range $9.8\leqq V_{\rm {LSR}}\leqq12.0$ km s$^{-1}$
derived from  (a) the C$^{18}$O ($J=3-2)$ emission line and (b) the N$_2$H$^{+}$ emission line.
\label{fig:v0}}
\end{center}
\end{figure}

To study the velocity structure of the clump, we made channel maps of some of the observed emission lines in Figure \ref{fig:channel}.
It can be seen that the lower velocity gas is located to the northwest and the higher velocity gas is located to the southeast.
To further investigate the velocity structure of the clump, we show mean velocity maps of C$^{18}$O($J=3-2$)\footnote{
The $V_0$ map for C$^{18}$O($J=3-2$) was measured from $\int{VT_{\rm mb}}\,dv/\int {T_{\rm mb}\,dv}$.}
and N$_2$H$^{+}$ (see Section \ref{sec:mass}) in Figure \ref{fig:v0}.
The maps show a clear velocity change near VLA1 in the GGD12-15 core, with a blue shift toward the northwest and a red shift toward the southeast.
This result is opposite to the directions of the blue and red lobes of the outflow, 
indicating that the ambient dense gas traced by C$^{18}$O and N$_2$H$^{+}$ is infalling.

\begin{figure}
\begin{center}
\includegraphics[scale=.4]{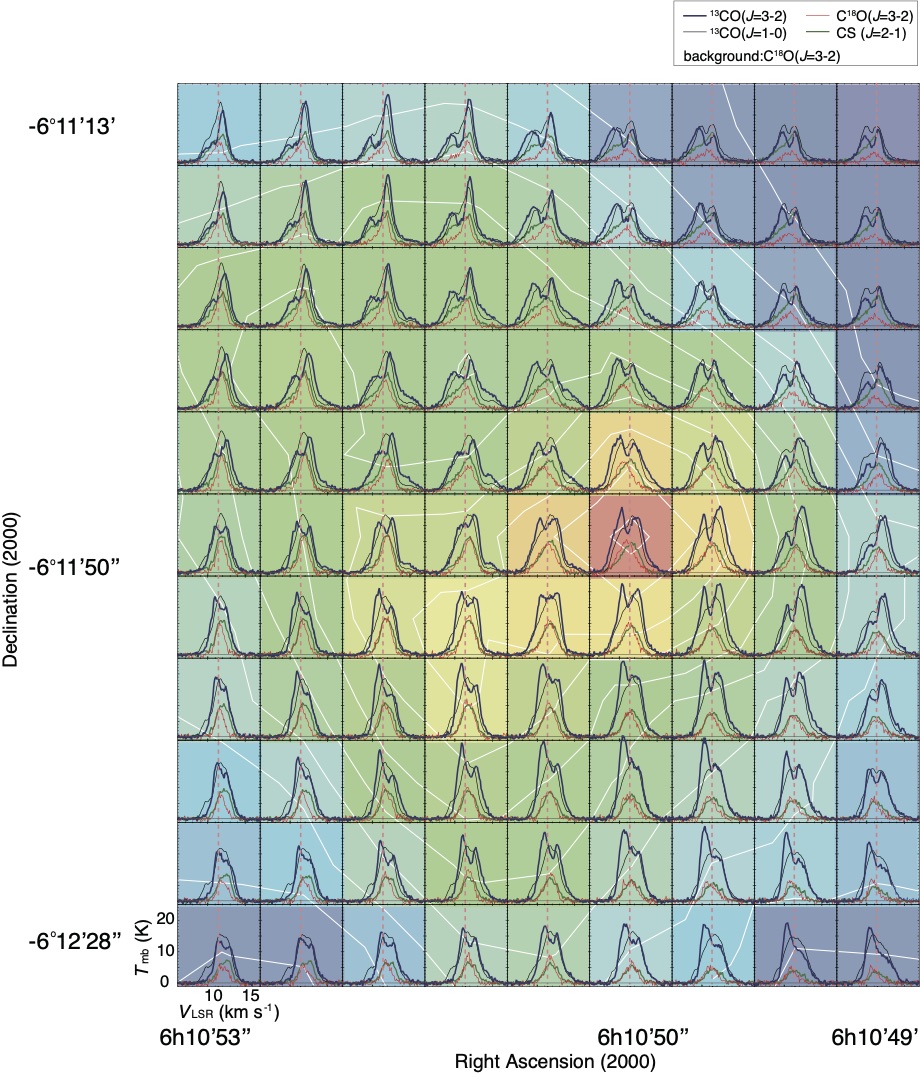}
\caption{
Spectral line profiles around VLA1. Blue, black, red, and green lines represent the $^{13}$CO ($J=3-2)$, $^{13}$CO ($J=1-0)$, C$^{18}$O ($J=3-2)$, and CS($J=2-1)$ emission lines, respectively. 
Each spectrum is displayed for each observed pixel in the C$^{18}$O ($J=3-2)$ intensity map shown in the background.
The vertical dashed line indicates the systemic velocity ($V_{\rm{LSR}}$ = 11.4 km s$^{-1}$).
\label{fig:promap}}
\end{center}
\end{figure}

In high-mass star-forming regions, blue asymmetrical profiles of optically thick lines, which is
suggestive of mass inflow, have been reported \citep[e.g.,][]{Reiter2011, He2016}.
We investigated line profiles of the $^{13}$CO ($J=3-2)$, C$^{18}$O ($J=3-2)$, $^{13}$CO ($J=1-0)$, and CS ($J=2-1)$ emission lines around the clump as shown in Figure \ref{fig:promap}. 
We also estimated $\tau$ of these lines\footnote{$\tau$ was derived as
$\tau=-{\rm{ln}}\left[{1-T_{\rm{mb}}}/{J(T_{\rm{ex}})-J(T_{\rm{bg}})}\right]$,
where $J(T)=T_{0}\diagup(e^{(T_{0} \diagup T)}-1)$ with $T_{0}=h\nu/k$. $T_{\rm bg}$ is the cosmic background temperature 2.73 K.}.
As expected, the line profiles of $^{13}$CO ($J=3-2)$, whose optical depth is typically $\gtrsim1$
around the center of the clump, show the double-peaked profiles with blue asymmetry
at several positions including at the position of VLA1. The dips in the spectra should be caused by the
absorption by colder gas in the foreground at the same velocities \cite[e.g.,][]{Zhou1992}.
On the other hand, the C$^{18}$O($J=3-2$) emission line appears more simple with
a single-peaked profile because the line is optically thinner ($\lesssim 0.3$).

\begin{figure}
\begin{center}
\includegraphics[scale=0.4]{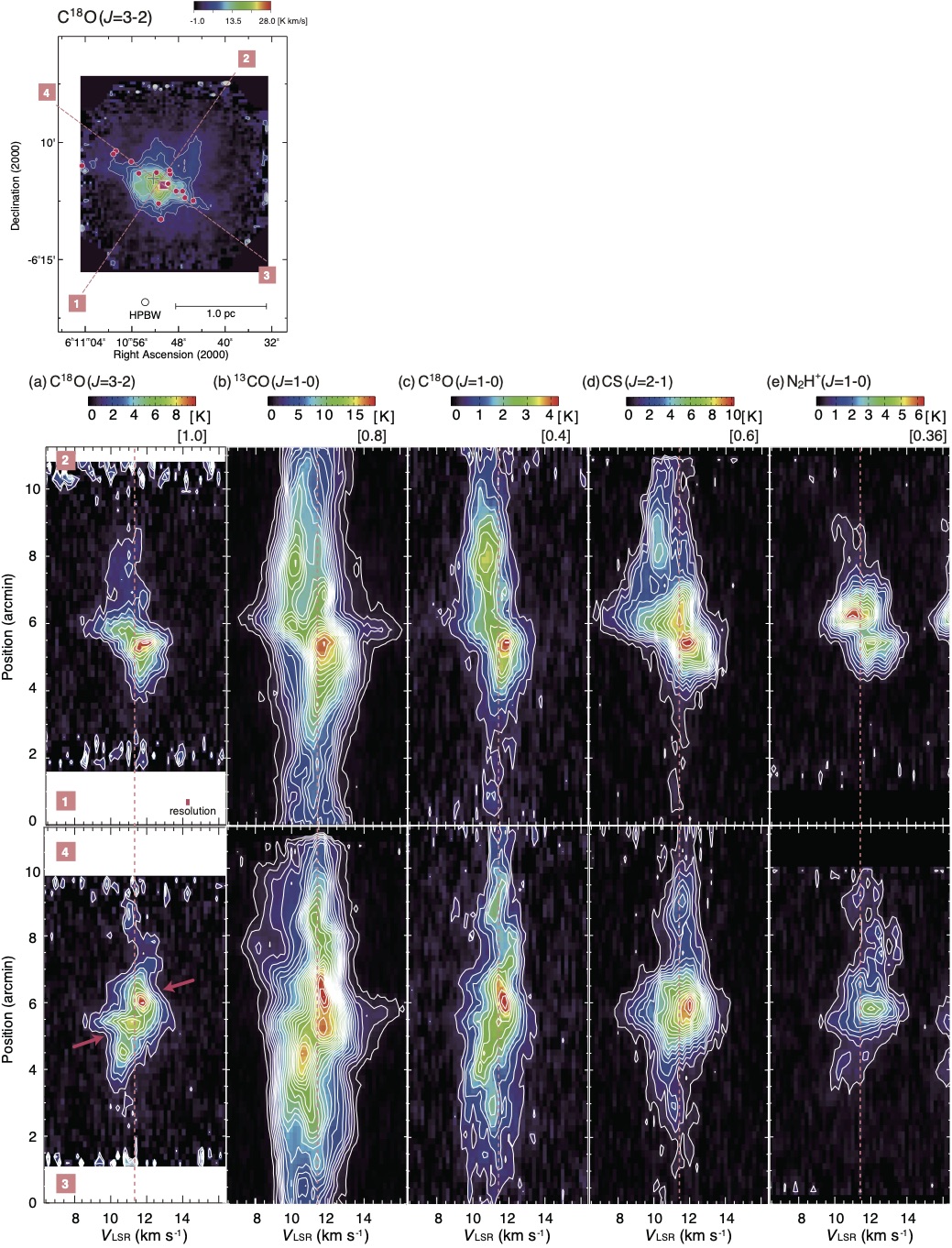}
\caption{
(a)--(e) PV diagrams of the observed emission lines. The upper panels are made along the cut $1-2$ shown in the C$^{18}$O ($J=3-2$) map at the top panel, and those of the lower panels are made along the cut $3-4$ set perpendicular to the $1-2$ line. 
The lowest contours and the contour intervals in units of K are indicated in the square brackets above each panel. 
The vertical dashed line indicates the systemic velocity ($V_{\rm{LSR}}$ = 11.4 km s$^{-1}$).
The two velocity components are indicated by the red arrows (see text).
\label{fig:all_pv}}
\end{center}
\end{figure}

Figure \ref{fig:all_pv} shows the PV diagrams of the observed emission lines.
The upper panels of (a)--(e) are made along the cut $1-2$ set parallel to the outflow, and those of the lower panels are made along the cut $3-4$ set perpendicular to the outflow. 
We point out two important features that can be seen in the figure.
One is shown in the upper panels of the figure (the cut $1-2$); there is blue-shifted emission with respect
to $V_{\rm{sys}}$ at the positions where the red lobe of the outflow is prominent (Figure \ref{fig:outflow2}), and in the same manner, there is red-shifted emission at the positions where the blue lobes of the outflow is prominent. 
In short, there are the velocity-shifted components (or velocity gradient) in the opposite sense to the outflow.
The other feature is shown in the lower panels of the figure (the cut $3-4$); we can recognize two well-defined peaks in C$^{18}$O($J=3-2$) around VLA1 (at $\sim6\pm1\arcmin$ in Figure \ref{fig:all_pv}, pointed by the red arrows), and a similar velocity gradient can be seen in other emission lines, except
for the CS and N$_2$H$^{+}$ lines whose critical density is much higher than the CO lines and therefore their PV diagrams show gas motion only of much denser regions compared to the CO lines.

Note that the PV diagrams in the figure are
measured along the cut $1-2$ set in the middle of the clump passing through the distributions of the Class I/0 objects
and along the cut $3-4$ set perpendicular to it, which are slightly different from the cuts centered
on VLA1 or the outflow-driving source IRS 9Mc.
This is because the PV diagrams appear more symmetrical with respect to the intersection of the two cuts,
suggesting that the intersection may be closer to the center of gravity of the clump rather than the position of VLA1 or IRS 9Mc.
We should note, however, that the basic features in the PV diagrams do not change significantly for the small change
in the positions of the cuts.

\section{Discussion} \label{sec:discussion}

Our previous statistical study of cluster-forming clumps found that a double-peaked feature in a PV diagram along the major axis
as seen in the lower panel of Figure 8(a) is often seen in molecular clumps that are in an early stage of cluster formation. 
The feature is very similar to those observed in the envelope of low-mass YSOs.
The results for the low-mass YSOs have been interpreted as an infalling envelope with rotation \citep{Ohashi1997}. 
With an analogue to the low-mass YSOs, we analyzed the massive cluster-forming clump S235AB in our earlier study and found that it can also be explained in the same way \citep{Shimoikura2016}.
In this paper, we made a simple model of an ellipsoidal clump with rotation and infalling motions similar to the case for low-mass YSOs \citep{Ohashi1997}, as was done for the S235AB clump, and tried to see if it could explain the observed velocity structure of the clump. 

\begin{figure}
\begin{center}
\includegraphics[scale=.4]{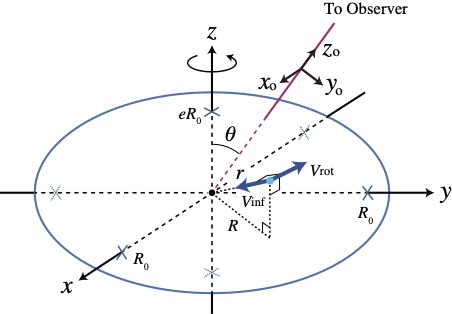}
\caption{A schematic illustration of the model taken from \cite{Shimoikura2016}.
\label{fig:model}}
\end{center}
\end{figure}

We assume an oblate model clump with an ellipticity $e_0$, and calculate PV diagrams along the major and minor axes of the clump to find parameters best-fitting the observed PV diagrams. We show a schematic illustration of the model in Figure \ref{fig:model}.
As illustrated in the figure, we assume that molecular hydrogen density $\rho$, infall velocity $V_{\rm inf}$, and rotation velocity $V_{\rm rot}$ can be expressed by the following equations,
\begin{equation}
\label{eq:density}
{{
\rho (r) = {\rho _0}{\left[ {1 + {{\left( {\frac{r}{{{R_{\rm d}}}}} \right)}^2}} \right]^{ - 0.75}} ~~,
}}
\end{equation} 
\begin{equation}
\label{eq:infall_velocity}
{{
{V_{\rm inf}}(r) = V_{\inf }^0\left( {\frac{r}{{{R_{\rm v}}}}} \right){\left[ {1 + {{\left( {\frac{r}{{{R_{\rm v}}}}} \right)}^2}} \right]^{ - 0.75}} ~~,
}}
\end{equation} 
and
\begin{equation}
\label{eq:rotation_velocity}
{{
{V_{\rm rot}}(R) = V_{\rm rot}^0\left( {\frac{R}{{{R_{\rm v}}}}} \right){\left[ {1 + {{\left( {\frac{R}{{{R_{\rm v}}}}} \right)}^2}} \right]^{ - 1}}
}}
\end{equation} 
where $\rho_{0}$, $V_{\inf }^0$, $V_{\rm rot}^0$, $R_{\rm d}$, and $R_{\rm v}$ are constants, and $r$ and $R$ are the distance from the center of the clump and from the rotation axis, respectively. 
We also assume that we are observing the clump at an angle of $\theta$ with respect to the rotation axis.

We selected the C$^{18}$O ($J=3-2$) emission line to compare with the model, because the PV diagrams of the line appear similar to the prediction of our model. The two basic features in the PV diagrams caused by the infall with rotation motions show up in PV diagrams derived from the other lines, but that of the C$^{18}$O ($J=3-2$) line appear more similar than the other lines.
This is probably because the line is optically thin and do not suffer from self-absorption in the spectra, unlike $^{13}$CO (Figure \ref{fig:promap}), as we assumed in our model. This is also probably because the line has a lower critical density and trace the total column density better than the other lines, such as CS($J=2-1$) or N$_2$H$^{+}$($J=1-0$), which may trace only the central dense regions in the clump (Figures \ref{fig:ii} and \ref{fig:all_pv}).

\begin{figure}
\begin{center}
\includegraphics[scale=.4]{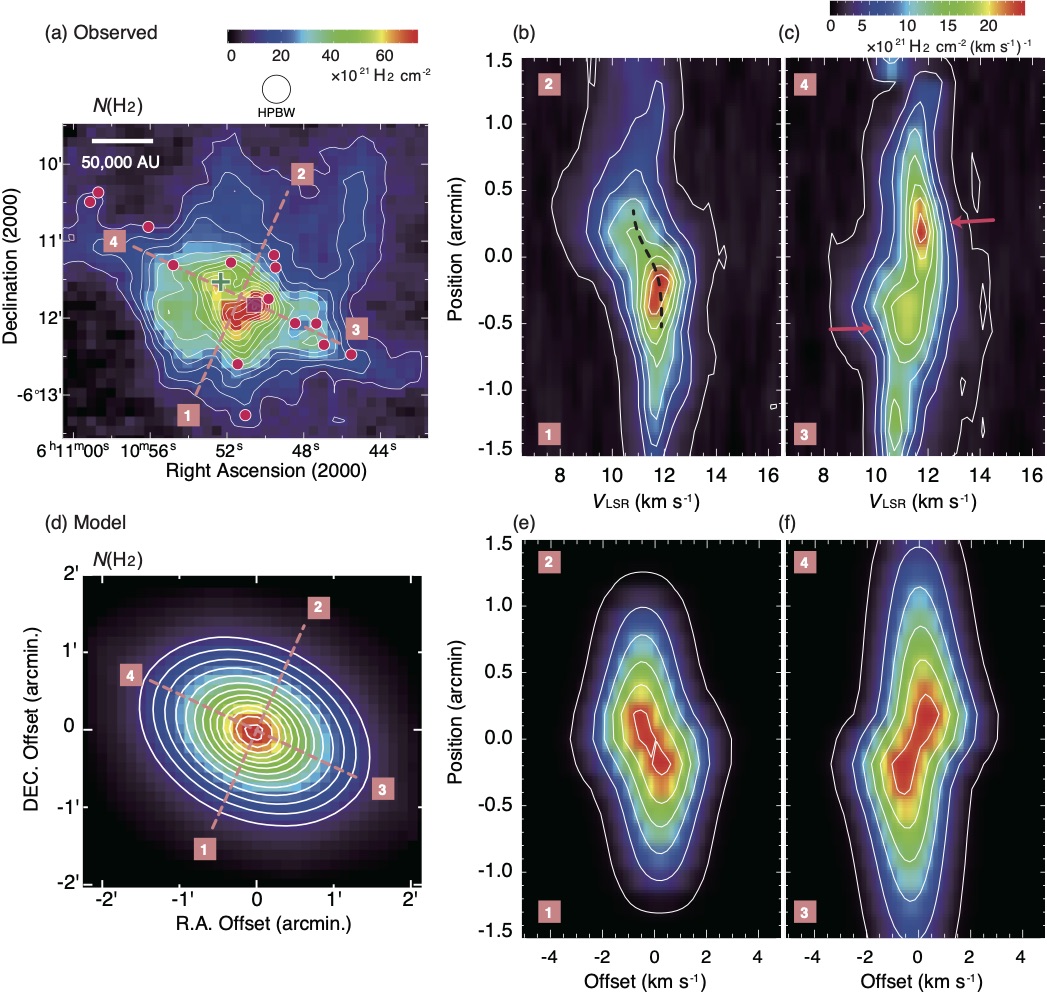}
\caption{
(a) Distributions of $N$(H$_2$) derived from the C$^{18}$O ($J=3-2)$ emission line.
(b) Observed PV diagrams measured along the minor axis (the cut $1–2$), and (c) that measured along the major axis (the cut $3-4$).
(d)--(f) Same as panels (a)--(c), but for the best-fit model.
The lowest contour and contour interval for both of the $N$(H$_2$) are $10\times10^{22}$ H$_2$ cm$^{-2}$ and $5\times10^{22}$ H$_2$ cm$^{-2}$, respectively. 
The lowest contour and contour interval for all of the PV diagrams are $4.4\times10^{21}$ H$_2$ cm$^{-2}$ (km s$^{-1}$)$^{-1}$, respectively. 
The dashed line in panel (b) shows the velocity gradient and the red arrows in panel (c) show the double-peaked feature (see text).
\label{fig:pv2}}
\end{center}
\end{figure}

In Figure \ref{fig:pv2}, we show the $N$(H$_2)$ map derived from the C$^{18}$O ($J=3-2$) line and the PV diagrams 
made along the same axes as in Figure \ref{fig:all_pv}.
To compare the observations with the model, we set the center of the clump on the plane of the sky at a position in the middle of IRS 9Mc and VLA1, and set the apparent minor axis (the cut 1-2 in Figure \ref{fig:pv2}(a)) of the clump parallel to the outflow axis, and set the major axis (the cut 3-4) orthogonal to it. 
The zero position of the outflow axis (in Figures \ref{fig:outflow1} and \ref{fig:outflow2}) is set at IRS 9Mc, and it is a little different from that of the rotation axis of the clump. 
As we mentioned earlier, this is because the PV diagrams of the clump in Figures \ref{fig:pv2}(b) and (c) appear more symmetrical, suggesting that the center of gravity of the clump may be slightly shifted to the southwest of IRS 9Mc (or the northeast of VLA1).

For the model,
we used the following observed parameters of the clump;
(i)  the typical line width $\Delta V$ = 2.0 km s$^{-1}$ (FWHM),
(ii)  rotation velocity $V_{\rm rot}^{\rm obs}$ = 0.2 km s$^{-1}$ at the equatorial radius of 69600 au (inferred from Figure \ref{fig:pv2}(c)),
(iii) apparent ellipticity = 0.71 obtained by fitting the $N$(H$_2$) distribution with a two-dimensional elliptical Gaussian function (Figure \ref{fig:pv2}(a)), and
(iv) the mass = 128.7 $M_\sun$ contained within the half-maximum contour of the $N$(H$_2)$ peak (Figure \ref{fig:pv2}).
The mass was estimated using the procedure described in Section \ref{sec:mass}, with the C$^{18}$O($J=3-2$) emission line instead of C$^{18}$O($J=1-0$). 
Details of the model are described in \cite{Shimoikura2016}.

Based on the model with varying $e_0$, $\theta$, and the other parameters in Equations (\ref{eq:density})-(\ref{eq:rotation_velocity}),
we calculated PV diagrams along the major and minor axes of the clump and used them to fit the observed PV diagrams shown in Figures \ref{fig:pv2}(b) and (c).
In the model, we calculated PV diagrams assuming that the clump is optically thin and did not incorporate the radiative transfer for simplicity.
The observed distribution of $N$(H$_2$) is rather complex, not a simple elliptical shape, as seen in Figure \ref{fig:pv2}(a). 
Thus, we did not compare it with the model, but we only compared the PV diagrams.

We searched for the best model parameters that minimize $\chi^2$. 
The best model is shown in Figure \ref{fig:pv2}(d)--(f).
Resulting parameters best fitting the observations are summarized in Table \ref{tab:model} with an error for the 90 $\%$ confidence level.
Though the reduced $\chi^2$ of the fit ($\sim3$) is rather poor due to the simplicity of the model,
two important features seen in the observed PV diagrams, i.e.,
the velocity gradient seen in Figure \ref{fig:pv2}(b) (delineated by the dashed line) and the two peaks in Figure \ref{fig:pv2}(c) (pointed by the arrows), are well reproduced by the model.

\begin{figure}
\begin{center}
\includegraphics[scale=.4]{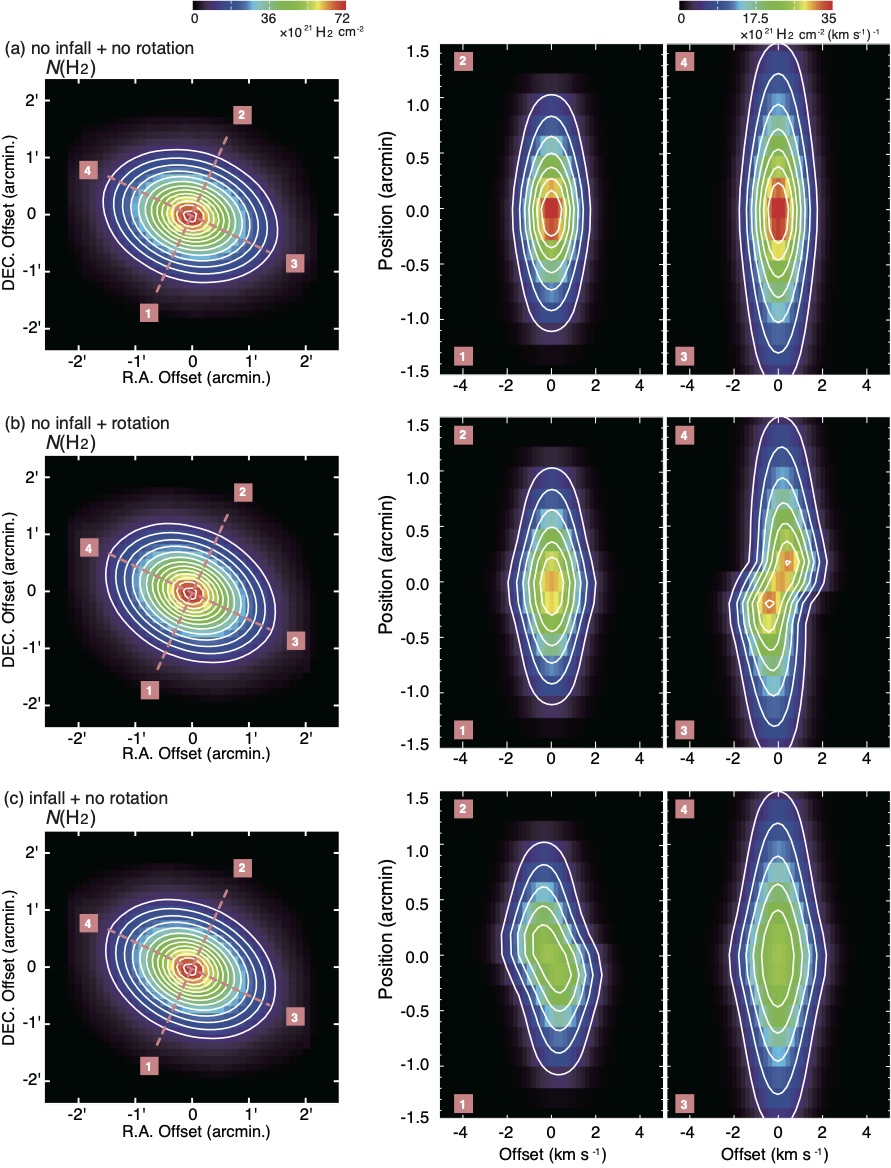}
\caption{
Model-based $N$(H$_2$) and its PV diagrams for the cases when the clump has (a) no infall nor rotation, (b) only rotation, and (c) only infall motions. 
The axes for creating the PV diagrams are the same as in Figure \ref{fig:pv2}.
The lowest contour and contour interval for the $N$(H$_2$) and the PV diagrams are the same as those used in Figure \ref{fig:pv2}, respectively.
\label{fig:model_PV}}
\end{center}
\end{figure}

The infall motion with rotation for an envelope of low-mass YSOs appears as the double-peaked feature in a PV diagram along the major axis \cite[e.g.,][see their Figure 8]{Ohashi1997}. 
We show model-based PV diagrams in Figure \ref{fig:model_PV} for the cases when the clump has (a) no infall nor rotation, (b) only rotation,
and (c) only infall motions. 
In Figure \ref{fig:model_PV}(b), the model only with rotation and no infall can also produce a similar double-peaked feature in its PV diagram along the major axis.
However, the results of model calculations with varying parameters showed that when infall motion is included in addition to the rotation, PV diagrams along the major axis tend to have a neck in the center, and the two peaks tend to appear more prominently, as seen in Figure \ref{fig:pv2}(f). 
Moreover, there is no velocity gradient in the PV diagram along the minor axis in this rotation-only model. 
In the same way, in the model only infall motion and no rotaion in Figure \ref{fig:model_PV}(c), there is a velocity gradient in the PV diagram along the minor axis, but the double-peaked feature in the PV diagram along the major axis is not seen. Thus, it is evident that none of the cases shown in Figure \ref{fig:model_PV} can reproduce the GGD12-15 observed PV diagrams. 
The observed two features in the PV diagrams can be reproduced only when the clump has both of the infall motion and rotation at the same time. 
These results strongly imply that the GGD 12-15 clump is collapsing with rotation.

Note that the best values of $R_{\rm d}=1.8\times 10^4$ au is much larger than $R_{\rm v}=5\times10^3$ au.
As we discussed in our previous study for S235AB \citep{Shimoikura2016}, we regard that the difference is probably
caused by the deficit of C$^{18}$O molecules in the dense central part of the clump due to the adsorption onto dust,
and the true $R_{\rm d}$ for gas (hydrogen) density should take a value close to $R_{\rm v}$.
Under the assumption of $R_{\rm d}=R_{\rm v}=5\times10^3$ au, we can estimate the mass infall rate from
the obtained model parameters (Table \ref{tab:model})
to be $\sim1.6\times10^{-3} M_\sun$ yr$^{-1}$ at the clump radius $1\times10^{5}$ au ($\simeq 0.5$ pc).
The mass loss rate by the outflow is $(2.9-6.4)\times10^{-4} M_\sun$ yr$^{-1}$.
This means that one third of the infalling mass is ejected by the outflow, which is consistent with a theoretical prediction \cite[e.g.,][]{Nakamura2014} and is also consistent with results of our previous study for S235AB \citep{Shimoikura2016}.

For some massive clumps showing a velocity structure similar to the GGD12-15 clump, a converging flow scenario has been proposed in which a molecular clump is formed by a large scale flow of H{$\,${\sc i}} gas, and at the same time the gas flow creates a dense core in which a cluster is formed \citep[e.g., SDC335;][]{Peretto2013}. 
In this case, observed velocity gradients are interpreted to be due to the gas flow towards the center of gravity.
\cite{Peretto2013} identified a core with a mass of 545 $M_\sun$ at the center of the massive IRDC, suggesting that the global collapse of the surrounding cloud caused the build-up of a large mass in the center.
The mass accretion in the clump is the same as \cite{Peretto2013} and the difference between their model and ours is that our model includes rotation in it.
Since the GGD12-15 clump shows velocity gradients along both the major and minor axes in the PV diagrams  in  Figures \ref{fig:pv2} (b) and (c), we conclude that both rotation and infall motopsn are necessary to explain the observations. 
In conclusion, we suggest that the GGD12-15 clump is undergoing gravitational contraction with rotation. 

\section{Summary} \label{sec:conclusions}

We observed the GGD12-15 region with several molecular emission lines, where cluster formation is occurring.
The main results of the observations can be summarized as follows.

\begin{enumerate}
\item
The analysis of the C$^{18}$O($J=1-0$) emission line shows that the molecular clump extends over $\sim2$ pc and has a mass of 2850 $M_\sun$ assuming a distance of 893 pc.
The C$^{18}$O($J=3-2$) emission line reveals an ellipsoidal structure centered on a compact {\Hii} region, 
which correlates well with the extent of the young cluster. 
Based on the N$_2$H$^{+}$ emission line, we identified a massive core forming the cluster in the center of the clump. 
The mass of the core is estimated to be $530M_\sun$ with a radius of 0.3 pc.

\item
There is a molecular outflow extending perpendicular to the distribution of C$^{18}$O($J=3-2$). 
The outflow consists of blue-shifted and red-shifted lobes that extend over $\sim$1 pc covering the entire clump, 
with an estimated mass of 11.7 $M_\sun$ and 10.6 $M_\sun$ for the blue and red lobes, respectively. 
The injection rate and dissipation rate of the turbulent momentum of the clump, $\dot{P}_{\rm lobe}$ and $\dot{P}_{\rm turb}$, were calculated, and we found $\dot{P}_{\rm lobe}> \dot{P}_{\rm turb}$.
However, we found that the clump is gravitationally bound and collapsing  by analyzing the entire velocity field of the clump, suggesting that the outflow only inject a marginal amount of turbulence inside the clump and has little effect on the dynamics of the clump.

\item
We found that two significant features appear in Position-Velocity (PV) diagrams of the observed emission lines.
One of them is the two well-defined peaks seen in the PV diagram taken along the major axis of the clump. 
The other feature is a velocity gradient in the opposite to that of the outflow seen in the PV diagram taken along the minor axis of the clump.

\item
We made a simple model of an oblate clump that gravitationally collapsing with rotation, and formulated the infall velocity, rotational velocity, and density distributions as a function of distance to the center of the clump. 
We found that the two features in the PV diagrams can be reproduced only when infalling and rotational motions are present, implying that the GGD 12-15 clump is collapsing with rotation.
Using the best-fit model parameters, the mass infall rate obtained from the total mass within the clump radius $1\times10^{5}$ au ($=\sim 0.5$ pc) is estimated to be $\sim1.6\times10^{-3} M_\sun$ yr$^{-1}$.
Comparing the value with a mass loss rate by the outflow, it is suggested that one third of the infalling mass is ejected by the outflow. 

\end{enumerate}

\begin{acknowledgments}

We are very grateful to the anonymous referee for providing useful comments and suggestions to improve this paper.
One of the authors, T.S., would like to thank Ms. A. Hirose, Ms. K. Shimizu, and Ms. T. Ishikawa for their help with the observations.
The Nobeyama 45-m radio telescope is operated by Nobeyama Radio Observatory, a branch of National Astronomical Observatory of Japan. The James Clerk Maxwell Telescope is operated by the East Asian Observatory on behalf of the National Astronomical Observatory of Japan; Academia Sinica Institute of Astronomy and Astrophysics; the Korea Astronomy and Space Science Institute; Center for Astronomical Mega-Science (as well as the National Key R$\&$D Program of China with No. 2017YFA0402700).
 Additional funding support is provided by the Science and Technology Facilities Council of the United Kingdom and participating universities and organizations in the United Kingdom and Canada.
The authors wish to recognize and acknowledge the very significant cultural role and reverence that the summit of Maunakea has always had within the indigenous Hawaiian community.  We are most fortunate to have the opportunity to conduct observations from this mountain.
\end{acknowledgments}

\appendix
\section{Distribution of the other molecular emission lines}\label{sec:appendix}
\begin{figure}
\begin{center}
\includegraphics[scale=.4]{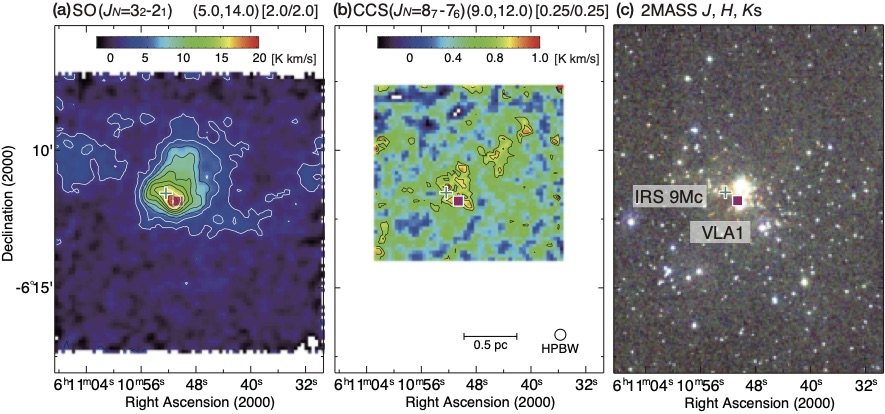}
\caption{
Same as Figure \ref{fig:ii}, but for the (a) SO and (b) CCS molecular emission lines.
(c) 2MASS $J$, $H$, $K_{\rm s}$ composite image around GGD12-15. 
\label{fig:ii2}}
\end{center}
\end{figure}

We show the integral intensity maps of the SO and CCS molecular emission lines in Figure \ref{fig:ii2}.



\begin{deluxetable*}{lcrcc} 
\tablecaption{The observed molecular lines \label{tab:line}} 
\tablehead{ 
 \colhead{Molecule} & \colhead{Transition}   & \colhead{Frequency}  &  \colhead{$\Delta T_{\rm mb}$} & \colhead{$\eta_{\rm mb}$}\\
 \colhead{} & \colhead{(GHz)}   & \colhead{(arcsec)} & \colhead{(K)}  & \colhead{} 
}
\startdata 
HC$_{3}$N	&	$J=10-9$	&	90.9790230 	&	0.16	&	0.47\\
N$_2$H$^{+}$	&	$J=1-0$	&	93.1737637 	&	0.18 &	0.47\\
CCS	&	$J_{N}$=8$_{7}-7_{6}$	&	93.8700980 	&	0.18& 0.47	\\
CS	&	$J=2-1$	&	97.9809530 	&	0.30 &	0.44\\
SO	&	$J_{N}$=2$_{3}-1_{2}$	&	99.2999050 	&	0.33 &0.44	\\
C$^{18}$O	&	$J=1-0$	&	109.7821760 	&	0.21 &	0.45\\
$^{13}$CO	&	$J=1-0$	&	110.2013540 	&	0.21 &	0.44\\
$^{12}$CO	&	$J=1-0$	&	115.2712040 	&	0.77 & 0.41	\\
C$^{18}$O	&	$J=3-2$	&	329.3305453 	&	0.64 &	0.64\\
$^{13}$CO	&	$J=3-2$	&	330.5879601 	&	0.48 &	0.64\\ 
\enddata 
\tablecomments{The rest frequency for the N$_2$H$^{+}$ line is taken from \cite{Pagani2009}, and those of the other lines are taken from the website of \cite{Lovas}.}

\end{deluxetable*}

\begin{deluxetable*}{crccccc} 
\tablecaption{Physical Parameters of the GGD12-15 clump measured in C$^{18}$O($J=1-0$) \label{tab:param1}} 
\tablehead{ 
\colhead{$\alpha$(J2000)} 	&	\colhead{$\delta$(J2000)}	&\colhead{${T_{\rm ex}}$}\tablenotemark{\rm*} &	\colhead{${V_{\rm LSR}}$}	&	\colhead{${\Delta V}$}		&\colhead{$\overline{\Delta V}$}	&	\colhead{$R$} 	\\
\colhead{($^h\:^m\:^s$)}	&	\colhead{(${\arcdeg}\:{\arcmin}\:\:{\arcsec}$)} &	\colhead{ (K)}	&	\colhead{(km s$^{-1}$)} 	&	\colhead{(km s$^{-1}$)} 	&	\colhead{(km s$^{-1}$)}	&	\colhead{(pc)}	
 }
\startdata 
06 10 51.0	&	-06 11 51	& 40.5 &	11.4	&	2.5	&	2.0	&	1.30 	\\
\enddata 
\tablecomments{\tablenotemark{\rm*}The parameter is derived from the $^{12}$CO data.}
\end{deluxetable*}

\begin{deluxetable*}{rrcccccc} 
\tablecaption{Pyshical Parameters of the GGD12-15 core measured in N$_2$H$^{+}$ \label{tab:param2}} 
\tablehead{ 
 	\colhead{$\alpha$(J2000)}	&	\colhead{$\delta$(J2000)}	&	\colhead{${T_{\rm ex}}$}	&	\colhead{${V_{\rm LSR}}$}	&	\colhead{${\tau_{\rm{tot}}}$}	&	\colhead{${\Delta V}$}	&	\colhead{$\overline{\Delta V}$}	&	\colhead{$R$}   	\\
   	\colhead{($^h\:^m\:^s$)}	&	\colhead{(${\arcdeg}\:{\arcmin}\:\:{\arcsec}$)} 	&	\colhead{ (K)}	&	\colhead{(km s$^{-1}$)} 	&		&	\colhead{(km s$^{-1}$)} 	&	\colhead{(km s$^{-1}$)} 	&	\colhead{(pc)}
 }
\startdata 
06 10 49.5	&	-06 11 21	&	20.9	&	10.9	&	1.4	&	1.6	&	1.6	&	0.3	\\
\enddata 
\end{deluxetable*}

\begin{deluxetable*}{lrrr} 
\tablecaption{Derived Mass \label{tab:mass}} 
\tablehead{ 
 \colhead{} 	&	\colhead{$M_{\rm LTE}$}	&	\colhead{$M_{\rm vir}$}	\\	
  \colhead{} 	&	\colhead{($M_{\sun}$)}	&	\colhead{($M_{\sun}$)}	
 }
\startdata 
the GGD 12-15 clump	&	2850 	&	1210 	\\
the GGD 12-15 core	&	530 	&	153	\\
\enddata 
\end{deluxetable*}

\begin{deluxetable*}{ccccccccc} 
\tablecaption{Outflow Parameters \label{tab:outflow}} 
\tablehead{ 
\colhead{lobe}	&	\colhead{$\vert V_{\rm sys}-V_{\rm max}\vert$}	&	\colhead{$S_{\rm lobe}$}	&	\colhead{$R_{\rm lobe}$}	&	\colhead{$t_{\rm d}$}	&	\colhead{$\tau_{\rm{CO}}$}	&	\colhead{$M_{\rm lobe}$\tablenotemark{\rm*}}	&	\colhead{$P_{\rm lobe}$\tablenotemark{\rm*}}	&	\colhead{$\dot{P}_{\rm lobe}$\tablenotemark{\rm*}}	\\
\colhead{}	&	\colhead{(kms$^{-1}$)} 	&	\colhead{(arcmin$^{2}$)}	&	\colhead{(pc)}	&	\colhead{(10$^{4}$yr)}	&	\colhead{}	&	\colhead{($M_{\sun}$)}		&	\colhead{($M_{\sun}$ km s$^{-1}$)}	&	\colhead{(10$^{-4}$$M_{\sun}$ km s$^{-1}$yr$^{-1}$)}
}
\startdata
blue&	14.0	&	12.4	&	1.2	&	8.9	&	3.6	&	$3.2-11.7$	&	$44.2-164.1$	&	$5.3-19.0$	\\
red	&	18.6	&	11.9	&	0.9	&	4.5	&	1.8	&	$4.9-10.6$	&	$91.5-197.1$	&	$19.3-42.0$	\\
\enddata 
\tablecomments{\tablenotemark{\rm*} Minimum and maximum estimates for parameters of the lobes, which are estimated using $\tau=0$
and $\tau={\tau_{\rm{CO}}}$, respectively.}
\end{deluxetable*}


\begin{deluxetable*}{ll} 
\tablecolumns{2} 
\tablecaption{Model Parameters best fitting the observed data\label{tab:model}} 
\tablewidth{0pt}  
\tablehead{ 
\colhead{Parameters}   &  \colhead{Values (Range for the 90$\%$ confidence level)} 
}
\startdata 
$e_0$	&	$0.2$	($0.16-0.49$)	\\
$\theta$ (deg)	&	$48$	($47-56$)	\\
$\rho{_0}$ (H$_2$ cm$^{-3}$)	&	$3.1\times10^{5}$	($2.0\times10^{4}-3.3\times10^{5}$)	\\
$V_{\rm inf}^0$ (km s$^{-1}$)	&	$2.4${\tablenotemark{\rm*}}($1.9-2.8$)	\\
$V_{\rm rot}^0$ (km s$^{-1}$)	&	$2.8${\tablenotemark{\rm*}}($2.8-4.7$)	\\
$R_{\rm v}$  (au)	&	$5\times10^{3}$	($0-2.0\times10^{4}$)	\\
$R_{\rm d}$  (au)	&	$1.8\times10^{4}$	 ($1.4\times10^{4}-2.3\times10^{4}$)	\\
\enddata 
\tablecomments{\tablenotemark{\rm*} The parameters are derived at $R_{\rm d}=R_{\rm v}=5\times10^{3}$ au (see text).}

\end{deluxetable*}



\end{document}